\documentclass{Interspeech}

 \interspeechcameraready

\title{End-to-End Diarization utilizing Attractor Deep Clustering}

\author[affiliation={1}]{David}{Palzer}
\author[affiliation={2}]{Matthew}{Maciejewski}
\author[affiliation={1}]{Eric}{Fosler-Lussier}

\affiliation{Computer Science and Engineering}{The Ohio State University}{U.S.A.}
\affiliation{Human Language Technology Center of Excellence}{The Johns Hopkins University}{U.S.A.}
\email{palzer.1@osu.edu, mmaciej2@jh.edu, fosler@cse.ohio-state.edu}

\usepackage{comment}
\usepackage{balance}

\begin{document}

\maketitle

\begin{abstract}
Speaker diarization remains challenging due to the need for structured speaker representations, efficient modeling, and robustness to varying conditions. We propose a performant, compact diarization framework that integrates conformer decoders, transformer-updated attractors, and a deep clustering style angle loss. Our approach refines speaker representations with an enhanced conformer structure, incorporating cross-attention to attractors and an additional convolution module. To enforce structured embeddings, we extend deep clustering by constructing label–attractor vectors, aligning their directional structure with audio embeddings. We also impose orthogonality constraints on active attractors for better speaker separation while suppressing non-active attractors to prevent false activations. Finally, a permutation invariant training binary cross-entropy loss refines speaker detection. Experiments show that our method achieves low diarization error while maintaining parameter count.
\end{abstract}

\section{Introduction}
\label{sec:intro}

Speaker diarization, the task of segmenting an audio stream into homogeneous speaker segments,
remains a challenging problem due to the inherent variability in acoustic conditions, the presence 
of overlapping speech, and the need to robustly handle a wide range of speakers \cite{anguera2012speaker,PARK2022}.
End-to-end neural diarization systems (EEND)  \cite{fujita2019end} allow for direct minimization of diarization errors during training.
The original EEND included an intermediate auxiliary loss of Deep Clustering (DPCL) \cite{hershey2016deep} to help discriminate speakers; DPCL itself was introduced as a method to explicitly create discriminable embedding vectors for an arbitrary number of classes and was evaluated on speech separation.

Further work in the EEND line led to Encoder-Decoder Attractors (EDA) \cite{horiguchi2020end} where an EDA module was responsible for encoding discriminable speaker representations directly from audio embeddings, allowing for a flexible number of speakers to be separated.  Neither this work nor subsequent EEND studies (e.g.,  
\cite{fujita2023neural,palzer2024improving}) utilized DPCL.

In this work, we propose a performant, compact diarization framework that synthesizes 
the strengths of conformer-based decoders, transformer-updated attractors, and an improved DPCL-style angle 
loss to promote robust, structured representations. By enhancing a conformer architecture with 
cross-attention to attractors and incorporating additional convolutional modules, we capture both 
local and global speaker information while keeping the network size manageable. 
We also reintroduce DPCL loss into the EEND framework (first seen in \cite{fujita2019end} but absent in subsequent EEND models) with a new extension
to enforce directional alignment between label–attractor vectors and 
audio embeddings, leveraging orthogonality constraints on active attractors to reduce cross-speaker 
confusion. Finally, we include a permutation invariant training (PIT) \cite{yudongpit} binary cross-entropy (BCE) loss 
term to refine speaker detection, effectively suppressing non-active attractors and minimizing 
false alarms.

\begin{figure}
    \centering
    \includegraphics[width=0.8\linewidth]{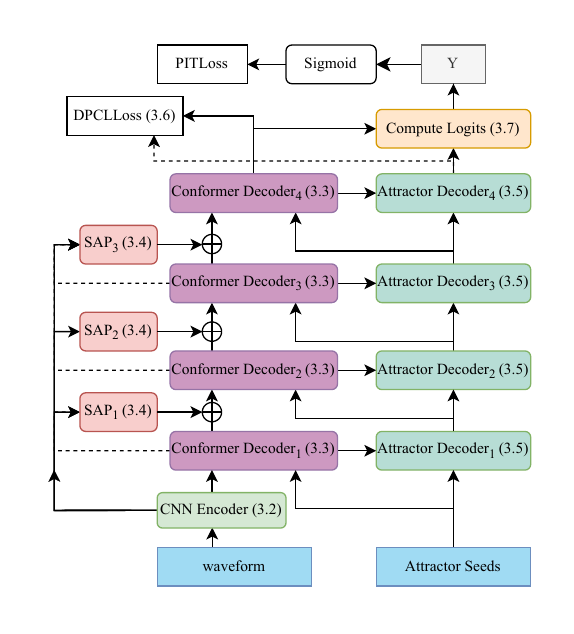}
    \caption{Example four layer EEND-CD architecture. Each block is described in subsections of Section~\ref{sec:method}.}
    \label{fig:model}
\end{figure}

Experiments demonstrate that our approach achieves competitive or superior 
diarization error rates (DER), while maintaining 
efficiency due to the compactness of the architecture. The key contributions of this paper are:
\begin{itemize}
    \item An \textbf{enhanced conformer decoder} design that integrates Latte 
          based attention \cite{dolga2024lattelatentattentionlinear} and an attractor-based cross-attention for refined speaker representation.
    \item An improved \textbf{DPCL-style angle loss extension} using label–attractor vectors to enforce 
          structured embedding alignments and orthogonality constraints.
    \item A novel \textbf{speaker detection refinement} strategy that explicitly works to reduce false attractor activations.
   
\end{itemize}

The rest of the paper is organized as follows. In Section~\ref{sec:related}, we discuss related 
studies in speaker diarization and representation learning. Our proposed method, including the 
conformer architecture, attractor updating mechanism, and extended DPCL loss, is described in 
Section~\ref{sec:method}. Section~\ref{sec:exps} details experimental setups and datasets, and 
Section~\ref{sec:results} presents our results, discussion, and future directions.

\section{Related Work}
\label{sec:related}

Speaker diarization involves partitioning an audio stream according to speaker identity, a task that has traditionally relied on statistical modeling and clustering approaches. Early methods often combined Gaussian Mixture Models (GMM) or i-vector \cite{dehak2011a} representations with clustering \cite{anguera2012speaker,sell2014speaker}, later transitioning to x-vectors \cite{snyder2018x} for more robust speaker embeddings. According to Park \textit{et al.}, \cite{PARK2022}, "many traditional speaker
 diarization systems, especially clustering-based systems, have
 only focused on non-overlapping regions and even the overlapping regions are excluded in the evaluation metric. (pg 22)"

Deep learning-based strategies have accelerated progress in speech and diarization \cite{PARK2022}, especially with the introduction of 
End-to-End Neural Diarization (EEND) \cite{fujita2019end}. DPCL projects time-frequency bins (or time frames) into an embedding space, then applies k-means to cluster an arbitrary number of speakers. In contrast, EEND directly predicts speaker activities in a permutation-invariant manner, but requires speaker counts a priori. More recently, attractor-based approaches \cite{horiguchi2020end,palzer2024improving,fujita2023neural} introduced learnable speaker prototypes (attractors) that iteratively refine speaker estimates, enabling the EEND models to adapt to an unknown or variable number of speakers. Despite these advances, challenges remain in capturing both local and global acoustic context efficiently, as well as maintaining scalability to longer recordings.

Work incorporating label dependency has been explored by \cite{fujita2023neural}, where intermediate attractors are utilized. These attractors are computed as a weighted sum of intermediate diarization predictions and subsequently added to each time frame before being processed by subsequent layers. This approach has been shown to reduce the missed speech rate, albeit at the cost of an increased false alarm rate.

The conformer architecture \cite{gulati2020conformer} has proven effective in speech processing tasks by combining local convolutions with global self-attention, improving upon pure transformer-based models in Automatic Speech Recognition (ASR). 
In diarization, conformer-based encoders have shown promise for generating robust speaker embeddings \cite{palzer2024improving,leung21_interspeech,liu21j_interspeech}. 
However, we are not aware of any diarization works that have explored deep conformer-based decoders that also integrate attractor-based cross-attention.

In parallel, various forms of optimized self-attention have been proposed to reduce quadratic complexity or to focus on latent representations of the sequence \cite{wang2020linformerselfattentionlinearcomplexity,deepseekai2024deepseekv3technicalreport}. Our work adopts a latent self-attention (Latte \cite{dolga2024lattelatentattentionlinear}) mechanism within each conformer decoder, providing an efficient means of capturing global context across frames while maintaining task performance.

Finally, angle-based losses, such as the DPCL-style angle loss, have been shown to encourage discriminable embeddings which show improvements in related fields such as speaker separation \cite{hershey2016deep, isik16single}. 
Following papers have pursued this idea further \cite{wang2018alternative} and offered alternative loss formulations.

\section{Proposed Method}
\label{sec:method}

In this section, we introduce our diarization framework, which combines a CNN-based feature extractor, 
conformer blocks with latent self-attention (Latte), attractor-based cross-attention, depthwise 
pooled residual connections, and a DPCL-style angle loss for structured embeddings. 
Figure~\ref{fig:model} illustrates the overall architecture.

\subsection{Overview of the Architecture}
Our framework takes a batch of recordings as input, where the recordings are processed into 23-dimensional mel features at a 10 ms frame rate. These frames are further stacked into windows of 15 frames using a hop of 10 frames such that each window overlaps with the next by 5 frames. We represent this input as a 4D tensor of shape \text{Batch} × \text{Time} × \text{Window} × \text{Mels}.
Previous models flatten these into a 3D tensor $\text{Batch} \times \text{Time} \times (\text{Window} \times \text{Mels})$, which is fed to the network directly, however we instead encode these into embeddings using a CNN described in the next section.  These encoded embeddings 
are then passed into a stack of conformer blocks where each block is paired with a 
transformer decoder that updates speaker attractors. Finally, we compute speaker logits using a binary cross entropy loss in a permutation-invariant manner.

\subsection{CNN-based Front End}
\label{sec:cnn_frontend}
The CNN treats each $\text{Window} \times \text{Mels}$ slice as a single “grayscale” image with one input channel, $\text{1} \times \text{Window} \times \text{Mels}$. We use a 5-layer CNN, each with a $\text{3} \times \text{3}$ kernel, except the last kernel of $\text{1} \times \text{2}$. We also increase the number of output filters (16, 32, 64, 128, 256) at each layer and downsample by half. This design preserves local time–frequency relationships and produces a robust representation before feeding into subsequent model components. By the final layer, 
the temporal dimension, $T$, remains intact (one embedding per time step), while the spatial dimensions 
collapse to a single $256$-dimensional feature vector, $X_t$. We then apply an RMSNorm to stabilize 
the embeddings before feeding them into the conformer blocks:
\begin{align}
    \text{X}_{t} \in \mathbb{R}^{256}, \quad 
    \forall \ t = 1, 2, \dots, T.
\end{align}
\subsection{Latte-based Conformer Decoders}
\label{sec:conformer}
Each conformer block follows the sequence:
\begin{enumerate}
    \item \textbf{Feed-Forward (FF)}: A positionwise two-layer MLP with a residual connection.
    \item \textbf{Latent Self-Attention (Latte)}: A multi-head latent-attention mechanism applied to 
    the time dimension \cite{dolga2024lattelatentattentionlinear}. 
    This captures global temporal relationships.
    \item \textbf{CNN}: A depthwise convolution module that captures local context.
    \item \textbf{Cross-Attention to Attractors}: The output of the CNN is used as queries, while 
    the current speaker attractors serve as keys and values (see Section~\ref{sec:attractors}).
    \item \textbf{CNN}: A second convolution block for further refinement.
    \item \textbf{Feed-Forward (FF)}: Closes the block with another FF sub-layer and residual.
\end{enumerate}
Each conformer block outputs $\text{Batch} \times \text{Time} \times 256$, which is then used to 
compute speaker logits or passed to subsequent layers (Section~\ref{sec:depth_pool}). This refinement strategy allows complex interactions between audio embedding and attractors.

\subsection{Self-Attentive Pooling (SAP) and Residual Injection}
\label{sec:depth_pool}
We maintain a list of conformer block outputs across network depth to be used in the self-attentive pooling (SAP) blocks. Suppose we have processed $d-1$ blocks, 
generating $\{\mathbf{X}^{(0)}, \dots, \mathbf{X}^{(d-1)}\}$, each of dimension 
$\text{Batch} \times \text{Time} \times 256$. Before feeding block $d$, we:
\begin{enumerate}
    \item Stack these embeddings along a depth axis:
    \begin{align}
        \mathbf{H}_{\text{stack}}^{(d-1)} 
        \in \mathbb{R}^{\text{Batch} \times \text{Time} \times (d) \times 256}.
    \end{align}
    \item Pool across this depth dimension using self-attentive pooling, yielding 
    $\mathbf{P}^{(d-1)} \in \mathbb{R}^{\text{Batch} \times \text{Time} \times 256}$.
    \item Residually add $\mathbf{P}$ to the current input 
    (the most recent audio embeddings):
    \begin{align}
        \mathbf{X}^{(d-1)} = \mathbf{X}^{(d-1)} + \mathbf{P}^{(d-1)}.
    \end{align}
    \item Feed $\mathbf{X}^{(d-1)}$ into the conformer block, $d$.
\end{enumerate}
This ensures each block sees a global summary of all prior representations, acting as a 
\emph{global residual} connection, and smooths training convergence.

\subsection{Attractor Updates via Transformer Decoder}
\label{sec:attractors}
We initialize a set of attractors $\mathbf{A} \in \mathbb{R}^{S \times E}$ as a set of $256$-dimensional 
vectors. These are \emph{tiled} across the batch as needed. After each conformer 
block computes its audio representation, a transformer decoder refines the attractors:
\begin{itemize}
    \item \textbf{Self-Attention:} Attractors attend to themselves to capture inter-speaker 
    relationships.
    \item \textbf{Cross-Attention:} The updated attractors then attend to the conformer output 
    for the current layer, enabling them to adapt to the updated audio representations.
\end{itemize}
Each attractor $\mathbf{a}_s$ thus evolves through the depth of the network, allowing multiple comparisons against it in the conformer. 

\subsection{Deep Clustering Auxiliary Loss}
\label{sec:dpcl_loss}
During initial experimentation we applied the original DPCL \cite{hershey2016deep}, however, we found that once a model had learned to detect speech and was learning to differentiate speakers it would diverge. This is most likely due to the mismatch between DPCLs orthogonality constraint and our BCE loss.

Our approach extends the original DPCL formulation \cite{hershey2016deep} in two ways; first, by leveraging multi-class diarization labels we recognize that constraints can be values other than parallel or orthogonal, and second, by allowing attractors themselves to specify the relative directions. These relative angles, in both cases, ensuring consistent geometric structure across training.

Let $\mathbf{X} \in \mathbb{R}^{T \times E}$ be the set of $L_2$-normalized audio embeddings for $T$ time frames, and $\mathbf{A} \in \mathbb{R}^{S \times E}$ be the set of attractors for $S$ speakers, where $E$ is the embedding dimension. The ground truth label matrix $\mathbf{Y} \in \{1, -1\}^{T \times S}$ indicates speaker activity at each time frame, with $1$ for active speakers and $-1$ for inactive speakers.

First, we propose MultiOpposite DPCL (MO-DPCL). For each time frame $t$, we construct a label vector:
\begin{align}
\mathbf{l}_t = \frac{y_{t}}{\max(||y_{t}||_2,\epsilon)}
\end{align}

The DPCL auxiliary loss is then computed using pairwise embedding distances:
\begin{align}
\mathcal{l}_{\text{DPCL}_{i,j}} = (\langle l_i, l_{j}\rangle - \langle\mathbf{x}_i, \mathbf{x}_{j}\rangle)^2,
\end{align}
where $\langle \cdot , \cdot \rangle$ denotes the inner product. This formulation ensures that embeddings which represent similar sets of speakers are clustered while pairs which differ will point away, saturating in opposite directions rather than orthogonality.

Additionally, we propose Attractor DPCL (A-DPCL). For each time frame $t$, we construct a label-attractor vector:
\begin{align}
\mathbf{l}_t = \frac{y_{t} \cdot \mathbf{A}}{\max(||y_{t} \cdot \mathbf{A}||_2,\epsilon)}
\end{align}

We compute the A-DPCL loss in a pairwise manner as above, and backpropagate though both audio embeddings, and label-attractors. While similar to MO-DPCL, this variant allows for more flexibility in the embedding geometry, and causes the relative geometry of the attractors and audio embeddings to be directly compared and reflected in each other.

These auxiliary losses are applied on the final audio embedding which are used in our logit calculation.

\subsection{Logits Computation and Orthogonal Attractor Loss}
\label{sec:logits_dpcl}
The last layer's attractors are split into a $256$-dimension final attractor, $a_s$, and a bias term, $b_s$.
We compute speaker logits by taking a dot product between the conformer output 
$\mathbf{x}^{(d)}_{t}$ and each attractor vector $\mathbf{a}_s$, adding both the attractor bias 
$b_s$ and a learned global bias $b_{\text{global}}$:
\begin{align}
    \mathrm{logit}_{t,s} 
    = \mathbf{x}^{(d)}_{t} \cdot \mathbf{a}_s + b_s + b_{\text{global}}.
\end{align}
We apply a sigmoid to obtain per-speaker probabilities, using a suppressive BCE loss that penalizes 
non-active attractors. Specifically, we push attractors that are not assigned to a speaker to zero exactly, as a clean signal, and only consider the bias terms in the prediction. Using a mean squared error (MSE) loss, we encourage 
orthogonality among \emph{active} attractors to reduce speaker confusion in overlapping speech.

\section{Experiments}
\label{sec:exps}

\subsection{Data Preparation}
Following the example of prior EEND studies \cite{fujita2023neural}, we prepare data from the 
CALLHOME and Switchboard-2 (Phase II, III) corpora. Unlike some earlier works, 
we do not have access to the Switchboard-2 (Phase I) or the NIST Speaker Recognition Evaluation 
data. Moreover, whereas \cite{fujita2023neural} restricts training to two-speaker mixtures, 
our system can handle a larger number of speakers. Therefore, we generate training mixtures 
from the Switchboard-2 (Phase II, III) dataset and the CALLHOME training set. Validation is 
performed on the CALLHOME validation split, and final evaluations use the CALLHOME test set.

\subsection{Acoustic Features}
Each audio segment is converted into 23-dimensional log-Mel filterbank features with a frame length 
of 10\,ms. We concatenate 15 consecutive frames (window size) with a 10\,ms hop, resulting in 
345-dimensional feature vectors that represent 100\,ms of audio. These frames serve as 
inputs to our 5-layer CNN front-end (Section~\ref{sec:cnn_frontend}). During training, we randomly 
select a 50\,s contiguous portion of each recording to accelerate training.

\subsection{Augmentation}
We employ noise augmentation from the MUSAN dataset \cite{musan2015}, injecting background 
audio to improve robustness. Following \cite{landini2022simulated}, we do not use any 
reverberation augmentation for simplicity. All audio is resampled to 8\,kHz prior to feature 
extraction.

\begin{table*}[ht]
    \centering
    \begin{tabular}{lc|c|c|c|c|c|c|c|c}
    \hline
    \multicolumn{2}{|c|}{Model} & \multicolumn{2}{c|}{Parameters} & \multicolumn{4}{c|}{DER (\%)} & \multicolumn{2}{c|}{SAD (\%)}\\
    \hline
     & & Total & Free & DER & MS & FA & CF & MS & FA \\
        1. EEND-EDA \cite{horiguchi2020end} &&6.4M&6.4M& 9.96 & 5.40 & 1.36 & 2.81 & 3.85 & 0.87\\
        2. EEND-EDA-deep \cite{fujita2023neural} &&29.3M&17.0M& 8.50 & 4.43 & 1.31 & 2.76 & 3.15 & 0.85\\
        3. EEND with Attribute Attractors \cite{palzer2024improving} (12 layers) &&33.7M&33.7M& 7.87 & 4.18 & 1.42 & 2.27 & 3.28 & 0.84\\
        4. EEND with Attribute Attractors + Conformer \cite{palzer2024improving} (12 layers) &&35.3M&35.3M& 6.98 & 3.63 & 1.95 & 1.41 & 3.68 & 0.82\\
        
        5. 7 + SAP + A-DPCL (final system, 5 layers) &&15.3M&15.3M& \textbf{4.99} & 3.60 & \textbf{1.12} & 0.27 & \textbf{2.45} & \textbf{0.47}\\
        \hline\hline
        \multicolumn{9}{c}{Ablation study on changes required to get to final system}\\
        \hline
        6. EEND with Conformer Decoder (12 layers)  && 22.2M &22.2M & 5.78 & 4.11 & 1.45 & \textbf{0.22} & 3.05 & 0.51 \\
        ~~~~~+ last layer Attractor Decoder + MO-DPCL &&&&&&&&\\
        7. 6 + per layer Attractor Decoder (reduce to 5 layers) && 14.8M &14.8M & 5.60 & \textbf{3.52} & 1.72 & 0.36 & 2.49 & 0.68 \\        
           \end{tabular}
    \caption{Diarization error rates (DER) for each model on the CALLHOME test set and component results.}
    \label{tab:results}
\end{table*}

\subsection{Training Configuration}
Our models are trained for 2000 epochs using AdamW optimization with a one-cycle learning rate 
scheduler. We use a batch size of 64, which balances efficiency and convergence stability. 
We select the best-performing checkpoint based on validation loss, 
then evaluate on the CALLHOME test set.

\subsection{Model Configuration}
For all experiments, the CNN front-end downsamples the 345-dimensional windowed features to 
256-dimensional embeddings, which are RMS-normalized before entering the stack of conformer blocks 
(Section~\ref{sec:conformer}). We use 5 layers of conformer blocks and attractor decoder blocks, all of hidden dimension $256$, 
each producing intermediate representations. The latent self-attention (Latte) mechanism 
operating within each conformer block has a hidden dimension of 128. Speaker 
attractors, initialized as $\mathbf{8} \times \mathbf{256}$ dimension vectors, are updated by a transformer decoder, enabling 
our system to track a large number of concurrent speakers.

\subsection{Losses and Evaluation}
We compute logits for each time frame and speaker attractor, combining a suppressive 
binary cross-entropy loss with a DPCL-style angle loss (Section~\ref{sec:logits_dpcl}). Evaluation focuses on standard diarization 
metrics diarization error rate (DER), and speech activity detection (SAD).

\section{Results and Discussion}
\label{sec:results}

\subsection{Evaluation on CALLHOME}

Table~\ref{tab:results} presents the diarization performance of our proposed system and several baseline models on the CALLHOME test set, measured in terms of diarization error rate (DER), missed speech (MS), false alarm (FA), and confusion (CF). Our final system, which integrates a conformer-based backbone, attractor-based cross-attention, and A-DPCL with self-attentive pooling, achieves a DER of 4.99\%, significantly outperforming existing methods.

Compared to the baseline EEND-EDA system \cite{horiguchi2020end} (DER 9.96\%) and the deeper EEND-EDA-deep model \cite{fujita2023neural} (DER 8.50\%), our approach reduces the overall error by nearly 50\%. Notably, our system  surpasses the recent EEND with Attribute Attractors (7.87\%) and its conformer-enhanced variant (6.98\%) by a substantial margin, demonstrating that our architectural innovations contribute meaningfully to diarization accuracy.

Looking at DER, our method achieves the lowest confusion rate (0.27\%), reflecting the effectiveness of our attractor-based decoding in maintaining clear speaker separation. While maintaining competitiveness in missed speech (3.60\%), we see a modest improvement in the false alarm rate (1.12\%) over previous systems. The final system also significantly improves in SAD missed speech (2.45\%) and false alarms (0.47\%), indicating robust detection of both speech and silence regions. 

\subsection{Original Model and Ablation Study}

Our original model, row 6 in Table~\ref{tab:results}, employed a 12-block conformer decoder integrated with one attractor-based decoder and MO-DPCL loss. This architecture allowed a multi-depth comparison between audio embeddings and static speaker embeddings leading to gains in speaker differentiation, achieving a DER of 5.78\% with the lowest confusion rate (0.22\%) across all evaluated systems. However, the model's relatively high missed speech (4.11\%) and false alarm rate (1.45\%) indicated potential inefficiencies in speaker detection.

Reducing the number of conformer blocks from 12 to 5 while incorporating a per-layer attractor decoder improves model size while seeing modest improvements in diarization error (DER 5.60\%). Introducing the SAP mechanism, for training efficiency, and A-DPCL loss further enhances speaker separation and reduces DER to 4.99\%. We do see a small increase in confusion, however the overall tradeoff is worth it. This analysis underscores the importance of each component, with the the conformer decoder effective at separating speakers, and A-DPCL loss particularly effective in lowering false alarms.

\subsection{Analysis and Future Directions}

The strong performance of our system can be attributed to several key design choices. The conformer backbone efficiently models local and global dependencies, while the attractor-based decoder ensures consistent speaker embeddings across layers. The addition of SAP facilitates the aggregation of temporal information, and A-DPCL loss enforces alignment between audio embeddings and speaker attractors,
resulting in reduced confusion and improved speaker differentiation.

Despite our strong performance, the gains over the previous EEND system in \cite{palzer2024improving} are primarily from decreases in false alarms and confusions; the miss rate remains largely unchanged.  We hypothesize that the the use of independent speaker attractors that are encouraged to have orthogonal representations allows the model to better represent regions where there previously were confusions, but a fuller analysis will be left to future work.

Future work will also focus on refining miss rates further and extending our approach to handle more complex multi-speaker environments with minimal computational overhead.

Our findings highlight the importance of balancing model complexity with performance and demonstrate that integrating structured embedding losses and attention mechanisms can significantly advance the state of the art in neural speaker diarization.


\bibliographystyle{IEEEtran}
\bibliography{template}

\end{document}